\documentstyle[12pt,epsf]{article}
\begin{document}
\hfill UDM-GPP-EXP-96-3\\
\begin{center}\Large
{\bf Dark Matter Search with Moderately Superheated Liquids}\\
\end{center}\normalsize
\begin{center}
L.A. Hamel, L. Lessard and V. Zacek\\
{\it Groupe de Physique des Particules, Universit\'e de Montr\'eal,
Canada}\\

and\\

Bhaskar Sur\\
{\it AECL, Chalk River Laboratories, Canada}\\
\end{center}

\begin{abstract}
We suggest the use of moderately superheated liquids in the form of
superheated droplet detectors for a new type of neutralino search experiment.
The advantage of this method for 
Dark Matter detection is, that the
detector material is cheap, readily available and that it is easily possible 
to fabricate a large mass detector. Moreover the detector can be made 
"background blind", i.e. exclusively sensitive to nuclear recoils.\\
\end{abstract}
\begin{center}
{\it 
Proceedings of the 2$^{nd}$ Workshop on The Dark Side of the Universe \\
Rome, Italy, Nov.1995}
\end{center}
\section{Introduction}
Current models explaining the evolution of the universe and the measured 
slight
anisotropy of the cosmic background radiation have in common, that they predict
an appreciable contribution of non-luminous, non-baryonic matter in the form of
a mixture of relativistic, light particles and non-relativistic, massive
particles (so-called Hot and Cold Dark Matter). Accelerator experiments and
results from the first round
of Dark Matter experiments explored up to now only a small range of masses and
types of possible candidate
particles. In particular they leave room for an interpretation of
Cold Dark Matter in terms of
weakly interacting massive particles with masses between 20 GeV and a
few hundred GeV. Here a
fitting candidate is the neutralino of the "minimal supersymmetric standard
model"(ref. [1]). The cross section of these particles are expected to be
of electroweak strength, with coherent or axial coupling. These particles
are
supposed to be concentrated in a self gravitating, spherical halo around
our galaxy with a
Maxwellian velocity distribution in the galactic frame with a mean velocity
of 300 km$/$sec and a local density at the solar system
of 0.3\, GeV/cm$^{3}$.
The detection reaction of neutralinos would be elastic scattering with a
detector
nucleus and the nuclear recoil energy would be the measurable quantity. Under
the given assumptions the mean recoil energy would be\\
\begin{equation}
<E_{R}> \approx 2\,keV M_{N}\,(GeV)\; [M_{X}/(M_{N} + M_{X})]^2
\end{equation}
where $M_{N}$ and $M_{X}$ are the masses of the detector nucleus and of the
neutralino respectively. The recoil spectrum falls off exponentially with
\begin{equation}
 dN/dE \approx \exp ( -E / <E_{R}> )
\end{equation}

For all detector materials the recoil energies are expected to be smaller
than 100 keV and depending on more detailed assumptions on the cross section
the expected event rates in the range between 4 to 20 keV are between 0.01
to 100 events/kgd. Therefore in order to ensure a reasonable countrate a
high target mass of more than 50 to 100 kg
is needed, especially if one wants to detect the 5\% annual variation in
count rate due to the motion of the earth around the sun. The latter
would be the decisive signature for the detection of Dark Matter candidates.
Due to background limitations, however, current experimental
sensitivities are still far away ($>$factor 100) to reach the small interaction
rates predicted. We propose a new approach for Cold
Dark Matter detection, which has the potential of substantially improved
sensitivity.

\section{Detection of Nuclear Recoils with Superheated Droplet Detectors}
Since a direct detection of Dark Matter candidates proceeds via elastic
scattering off a detector nucleus one has to rely entirely on the observation 
of the small ionizing
signal of the recoiling nucleus in the energy range of several keV to several
tens of keV. Since present detectors are sensitive to all kind of ionizing
radiation, an extremely low level of radioactive impurities in the detector
material and its surroundings is necessary, as well as powerfull active and
passive background rejection techniques. In order to reduce the overall
background sensitivity we suggest to apply a detection method,
which is exclusively sensitive to the high ionization density of recoiling
nuclei with A $>$10, but is insensitive to the much smaller specific ionization
of ordinary $\alpha, \beta, \gamma$-radiation. As is well known from bubble
chamber operation, vapor droplet formation in superheated liquids is a process
of precisely this kind. But in contrast to the usual cyclic bubble chamber
operation, the detection of nuclear recoils requires only moderately
superheated liquids and a quasi-continuous operation becomes
possible (ref.[2,3]).
  It turns out that this technique is successfully applied in
superheated drop detectors for neutron dosimetry. Here 10 to 20$\mu$m
diameter droplets of a superheated
liquid e.g. CCl$_{2}$F$_{2}$ (Freon 12) are dispersed homogeneousely in
an elastic, clear medium, such as water saturated polyacrylamide gel
(ref. [4,5,6]). Each of these droplets acts as an individual miniature 
bubble chamber. Under
ambient temperature and pressure, i.e. 25$^{0}$C and 1 bar, the droplets remain
in a metastable,
superheated condition. Only when a neutron hits a droplet the recoiling nucleus
triggers the explosive formation of a gas bubble about 1 mm in diameter.
This event is
accompanied by an acoustic shock wave, which can be detected with piezoelectric
transducers.
Neutron counters of this kind are commercially available and have neutron
detection
thresholds as low as 10 keV, i.e. precisely in the
range of sensitivity needed for our application. They are insensitive
to $\beta$- and $\gamma$- radiation, as can be shown by placing the detectors
close to a strong $^{60}$Co source. The devices used in our particular tests
were obtained
from Bubble Technology Industries (ref.[7]). The polymer is contained in
a transparent polycarbonate test tube with an active detection volume of
4 cm in length and 1.4 cm in diameter. The device is equipped with a
piston. Upon unscrewing the piston, the detector becomes sensitive. After
an exposure the bubbles are held in place by the gel until the piston is 
reset and the gas bubbles are
again reduced to liquid droplets. Properly recompressed after every
exposure, the detectors can be used over years. Standard detectors for 
dosimetry are
loaded with about 0.4\% active material, with typical sensitivities of
about 20 bubbles/mrem for neutrons above 100 keV. Upon our request
detectors were fabricated with 3\%, 10\% and 25\% loading. Up to 
40\% loading appears feasible with the existing technology.

\section{Event Detection and Detector Performance}
Our experimental efforts so far concentrated on the following issues:
event detection, localization, detector performance and background studies.
In detectors with more 
than 3\% loading the elasic medium becomes opalescent, non-transparent and
eventually optical detection of the bubbles becomes impossible. For this 
reason and also in order to be able to trigger electronically on the events
we investigated acoustic detection of bubble formation. Piezoelectric sensors 
had already been shown 
to be sensitive to the sound of a vaporizing droplet (ref.[8]).\\
 
\begin{figure}[h]
\epsfysize=10.00cm
\epsfxsize=12.00cm
\begin{center}
\epsffile{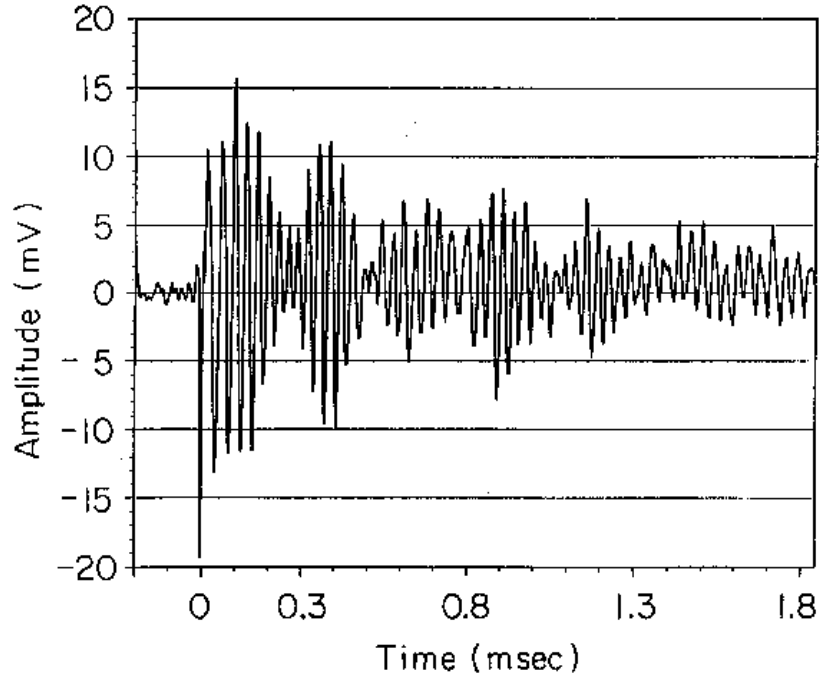}
\end{center}
\caption{\footnotesize{A typical signal from explosive 
droplet vaporization following
neutron interaction in a superheated droplet detector. The event starts 
off with a sharp negative pulse of about 2$\mu$sec duration and lasts
up to several milliseconds}}
\end{figure}

In order 
to simulate a larger detector volume several BTI test tube devices were 
immersed in a (28x28x28) cm$^{3}$ water tank, with four piezoelectric sensors 
installed at the tank walls. The piezoelectric detector surface makes
direct contact with 
the water. The sensors are wideband detectors (CANDED type-WD)
with a sensitivity band width ranging from 100 to 1000 kHz. Their signals 
are fed into low noise current
sensitive preamplifiers LM6365 with a gain of 15V/$\mu$A and a noise 
characteristic of 
1.5 pA/$\sqrt{Hz}$. At 5 cm distance a bubble event gives a signal
with a maximum amplitude between 50 and 200 mV above a noise band of 2 mV rms 
at the preamp output. The signals from bubble
formation are very characteristic (see Fig.1) and easily distinguishable 
from background
signals: all bubble signals start off with a sharp negative pulse lasting about 
2$\mu$sec; after this the piezoelectric sensor starts heavily ringing at a 
frequency of about 25 kHz. The amplitudes of these rather fast 
oscillations are then modulated by a much slower (kHz) oscillation which varies 
from event to event. The whole phenomenon lasts several msec. Our tests 
indicate that acoustic time-of flight methods can be used to locate bubbles.
For these measurements the 
preamp signals are fed into a FADC system and the event position is calculated
using a neural network trained for our detector geomtry. A video recording 
allows correlating the
electronic signals to individual bubbles in the (low loading) detectors. 
Our tests indicate that an event localization is
possible with a resolution of 0.5 cm in the tank. In a different experiment 
we measured the speed of sound in a detector gel with 0.4\% loading 
to be 1200$\pm$90 msec$^{-1}$, close to the speed of sound in water 
(1440 m$s^{-1}$).

In order to study the detector performance we spiked several detectors of 
different loading (0.4\%, 0.8\%, 1.5\%, 3.0\%) with a known (1-10 Bq) 
$\alpha$-activity of $^{241}$Am.
The 100 keV Am recoils come in coincidence with a 59.4 keV $\gamma$-ray, 
which allows in principal
the determination of recoil detection efficiencies. A variation of temperature 
translates into a variation of superheat according to the phase diagram
and therefore by varying the operating temperature of the 
detector we can study its response to different ionization densities. 
In particular we find a step-function like increase in countrate between 
37 and 40$^{0}$C where the 
detector becomes sensitive to $\alpha$-particles. By going to still higher 
temperature a strong rise in countrate sets in above 45$^{0}$C, where 
electrons and $\gamma$-rays start to trigger the detector.

\section{Background Considerations}
Being not sensitive to conventional 
$\alpha, \beta$ and $\gamma$ - radiation is the big asset of our detector.
Still the superheated droplets will be sensitive to the 100 keV nuclear
recoils following 5 to 6 MeV $\alpha$- decays due to the presence of U/Th
daughters in the detector material. This background can be controlled in two
ways: by raising the threshold of the detector, i.e. by varying the operating
parameters of pressure or temperature, the $\alpha$- induced background can
be measured separately and subtracted; alternatively a sensitivity of
the order of 10$^{-2} $cts$/$kg$/$day can be reached if the $^{235}$U
and $^{232}$Th
contamintion is brought down to a level of of 10$^{-14}$g$/$g. Since
our detector consists essentially of water (58\%), Freon (40\%) and about
2\% acrylomide we
are confident that this low level of activity can be achieved. Highly purified
water of the SNO-collaboration is available with a purity as high as
2 to 5x10$^{-15}$g$/$g U/Th.
  
In order to assey the present intrinsic background and to understand its 
origin, measurements were performed on the surface and at the location 
of the Sudbury Neutrino Observatory (SNO) 2000 m under ground with and 
without 20 cm 
water shielding. At the surface 0.5 cts/d were 
recorded for standard detectors with 0.4\% loading; in the mine two detectors 
were exposed up to now and showed no counts after 80 days. The detector masses 
envolved in these 
measurements were however too small to allow meaningful conclusions about the
intrinsic background of the devices. Independent of these direct measurements
the detector material was also tested for radioactive contaminations at the 
Gran Sasso low activity counting facility. Limits of 2.2 Bq/kg
were obtained for the $^{232}$Th contamination and again the measurements were 
compromised by the small mass of the test samples. On the other hand  
substantial $^{134}$Cs, $^{137}$Cs activities of about 0.2Bq/kg were found 
in the samples (this is due 
to the presence of CsCl salt, which is mixed into the gel in order to match 
the gel density to the  density of liquid Freon). Although the associated
$\beta$- and $\gamma$ activities are not harmfull by themselves for our 
application, we suspect an
important contamination of U/Th of the unpurified salt itself. This point 
is subject to further clarification.

Since we are handicapped by the small size of our test counters, we foresee 
as a next step to build a 1~kg prototype for more detailed
background studies on surface and underground. It will be read 
out with six piezoelectric sensors. This detector will later be installed at 
the SNO site, where space is already foreseen for this experiment.  
With the detector installed in a deep underground laboratory (SNO),
consideration has still to be given to the fast neutron component coming
from the rock walls. With the measured flux of 3 x 10$^{-6}$ n/cm$^{2}$sec
a passive shielding of 1m (borated) water will reduce the neutron
induced countrate to a level of sev. 10$^{-3}$n/day.

\section{Conclusions}
We suggest the use of moderately superheated liquids in the form of
a superheated droplet detector for a new type of dark matter search experiment.
The advantage of this method for 
Dark Matter detection is, that the
detector material is cheap, readily available and that it is easily possible 
to fabricate a large mass detector.
From its easy operating conditions and 
its suitable isotopic composition ($^{19}F$ is a spin-$1/2$$^{+}$ isotope) 
CCl$_{2}$F$_{2}$, i.e. Freon 12, is an interesting active material. Even more 
attractive because of its higher mass is CF$_{3}$Br. It has a similar vapor 
pressure curve as Freon 12 and is non-inflammable. Compared to alternative  
techniques our method is insensitive to $\alpha$, $\beta$ and $\gamma$ 
radiation, and avoids the need of complex cryogenics.

\section{Acknowledgement}
We thank M. Beaulieu, J.-P. Martin and G. Beaudoin (UdeM) for their 
assistance in designing
our readout system and for setting up the neural net. We are indepted 
to R. Bernabei and the Gran Sasso
low level counting team for their valuable help and we thank the BTI 
scientific staff for fruitful cooperation.

\section{References}
ref. [1]: A. Bottino et al.; Phys. Lett. B295 (1992) 330\\
ref. [2]: V. Zacek, Il Nuovo Cimento, 107A (1994) 291\\
ref. [3]: G. Riepe and B. Hahn, Helv. phys. Acta, 34 (1961) 865\\
ref. [4]: R. Apfel, Nucl. Inst. and Meth. 162 (1979) 603\\
ref. [5]: H. Ing and B.C. Bimboim, Nucl. Tracks 8 (1984) 285\\
ref. [6]: J. Sawicki, Nucl. Inst. and Methods. A 336 (1993) 215\\
ref. [7]: Bubble Technology Industries Inc., Chalk River, Canada KOJ1JO\\
ref. [8]: Apfel and S.C. Roy, Rev. Sci. Instrum. 54 (1983) 1397\\
\end{document}